\documentclass[twoside,fleqn]{ActaStyle}
\usepackage{times,cite}
\usepackage{here}
\usepackage{amssymb}

\usepackage{lscape}             % if you need landscape enviroinment
\usepackage{graphicx,epsfig}    % if you need to include figures in PostScript
\usepackage[tbtags]{amsmath}    % if you need more math

\def\authorheading{R.~Czy{\.z}ykiewicz, P.~Moskal et al.}

\def\shorttitle{The analysing power for the $\vec{p}p\to pp\eta$ reaction at Q=10~MeV}

\begin{document}
%% ---------------------------------------------------

%%%  page range, first and last page
\pagerange{1}{10}

\title{
THE ANALYZING POWER 
FOR THE $\vec{p}p\to pp\eta$ REACTION 
AT Q=10~MeV
}

\author{
R.~Czy\.zykiewicz\email{r.czyzykiewicz@fz-juelich.de}$^{a,b}$, P.~Moskal$^a$, H.-H.~Adam$^c$, A.~Budzanowski$^d$, 
E.~Czerwi\'nski$^a$, D.~Gil$^a$, D.~Grzonka$^b$, M.~Janusz$^{a}$, L.~Jarczyk$^a$, B.~Kamys$^a$, A.~Khoukaz$^c$, K.~Kilian$^b$, P.~Klaja$^a$, 
B.~Lorentz$^b$, J.~Majewski$^{a,b}$, W.~Oelert$^b$, C.~Piskor-Ignatowicz$^a$, 
J.~Przerwa$^a$, J.~Ritman$^b$, H.~Rohdjess$^e$, T.~Ro\.zek$^{b,f}$, R.~Santo$^c$, T.~Sefzick$^b$, M.~Siemaszko$^f$,
J.~Smyrski$^a$, A.~T\"aschner$^c$, K.~Ulbrich$^e$, P.~Winter$^{b\dagger}$, M.~Wolke$^b$, 
P.~W\"ustner$^g$, Z.~Zhang$^b$, W.~Zipper$^f$
}
{
$^a$Institute of Physics, Jagellonian University, 30-059 Cracow, Poland \\
$^b$IKP, Forschungszentrum J\"ulich, 52425 J\"ulich, Germany \\
$^c$IKP, Westf\"alische Wilhelms-Universit\"at, 48149 M\"unster, Germany \\
$^d$Institute of Nuclear Physics, 31-342 Cracow, Poland \\ 
$^e$Institut f\"ur Strahlen- und Kernphysik, Rheinische Friedrich-Wilhelms-Universit\"at, 53115 Bonn, Germany \\
$^f$Institute of Physics, University of Silesia, 40-007 Katowice, Poland \\
$^g$ZEL, Forschungszentrum J\"ulich, 52425 J\"ulich, Germany \\
$^\dagger$present address: Paul Scherrer Institute, 5232 Villigen, Switzerland \\ 
}

%%% Date of submition
\day{November 28, 2005}

%%% abstract of the paper
\abstract{
The analyzing power A$_y$ for the $\vec{p}p\to pp\eta$
reaction has been determined 
at the beam momentum $p_{beam}~=~2010$~MeV/c, corresponding
to the excess energy Q~=~10~MeV. In the paper the method of 
the data analysis is briefly presented.
%The preliminary experimental results 
%%are bared with 
%%rather large uncertainties,  
%compared with the predictions of pseudoscalar and vector meson 
%exchange models favor slightly the description 
%%Preliminary experimental data, encountered with
%%the predictions of the pseudoscalar- and vector meson exchange models
%%tend to slightly follow the description 
%in which the pseudoscalar exchange in the excitation 
%of the $S_{11}(1535)$ resonance plays the most important role.
}

%%% PASC numbers of your article
\pacs{
14.40.-n, 13.60.Le, 14.40.Aq 
}

% %%%%%%%%%%%%%%%%%%%%%%%%%%%%%%%%%%%%%%%%%%%%%%%%%%%%%%%%%%%%%%%%%
\section{Introduction}
% %%%%%%%%%%%%%%%%%%%%%%%%%%%%%%%%%%%%%%%%%%%%%%%%%%%%%%%%%%%%%%%%%
\label{sec:intr} \setcounter{section}{1}\setcounter{equation}{0}

\begin{figure}[H]
\begin{minipage}[t]{0.48\linewidth}
\begin{center}
\includegraphics[width=2.5in]{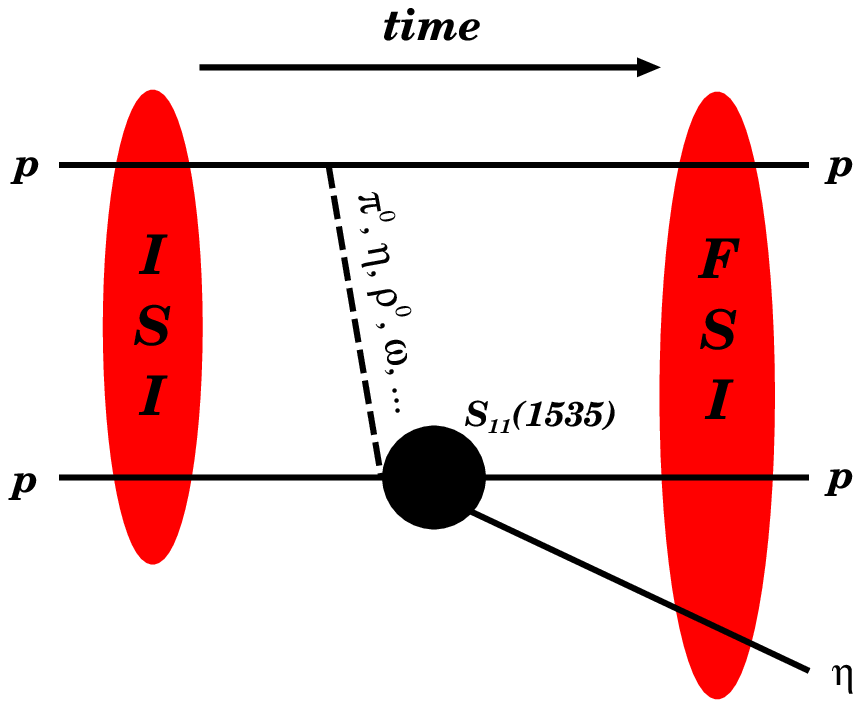} \\
\end{center}
\caption{
The Feynman diagram for one of the possible explanations of the 
$\eta$ meson production in elementary proton-proton collisions. 
ISI and FSI stand for the initial and final state interactions, 
respectively.  
}
\label{S11}
\end{minipage}%
\hspace{0.04\textwidth}%
\begin{minipage}[t]{0.48\linewidth}
\begin{center}
\includegraphics[width=2.5in]{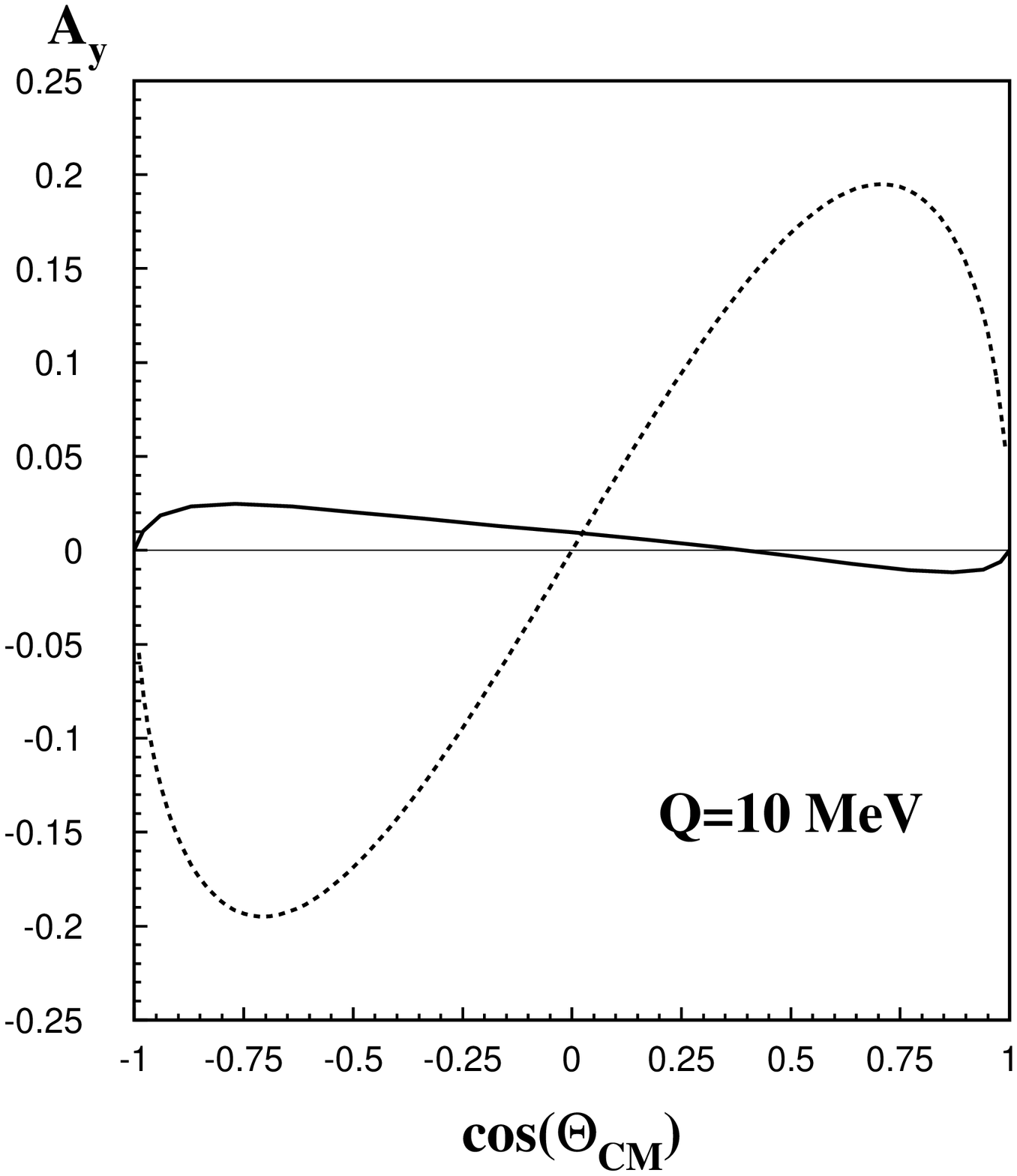} \\
\end{center}
 \caption{ Predictions of the analyzing power for the $\vec{p}p\to pp\eta$ reaction
at Q=10~MeV as a function of the center-of-mass (CM) polar angle of $\eta$ ($\theta_{CM}$). 
Full line are the predictions of the pseudoscalar meson exchange model~\cite{nakayama}, 
whereas the dotted line represents the results of the calculations, based on the vector meson exchange~\cite{wilkin}. 
}\label{modele}
\end{minipage}
\end{figure}

The close-to-threshold measurements of the total cross section 
for the $pp\to pp\eta$ reaction~\cite{bergdolt,chiavassa,
calen1,calen2,hibou,smyrski}, the measurements of the differential
cross sections~\cite{calen3,tatischeff,abdel}, the high statistic
investigations of the pp$\eta$ final state dynamics~\cite{moskal,hab} as well as 
the results of the first ever analyzing power experiment
for the $\vec{p}p\to pp\eta$ reaction~\cite{winter} are in  
agreement with the hypothesis that the main contribution 
to the production of the $\eta$ meson in the proton-proton
collisions comes from the mesonic excitation of the S$_{11}$(1535)
resonance and it's further decay into the p$\eta$ pair~\cite{wilkin,hanhart}, as depicted
in Fig.~\ref{S11}.   
However, 
%there is still little known about the production 
%mechanism of the $\eta$ meson in the elementary hadronic collisions,
%in particular, 
it is still not settled what are the relative contributions
to the production mechanism originating from each meson exchange 
denoted in Fig.~\ref{S11}. A possible way 
for elucidation of this problem is a precise measurement of 
the polarization observables, as they are very sensitive to the 
model assumptions on the type of the meson being exchanged in order
to excite the S$_{11}$ resonance. For example, from Fig.~\ref{modele}, where the 
predictions of the 
analyzing power for the $\vec{p}p\to pp\eta$ reaction at the excess energy
Q~=~10~MeV are shown, it can be noticed that the analyzing power function strongly
depends on whether we deal with pseudoscalar~\cite{nakayama} (full line)
or vector meson exchange model~\cite{wilkin} (dotted line). 

Therefore, in order to verify the validity of these two 
models, an experiment devoted to a determination of the 
analyzing power for the $\vec{p}p\to pp\eta$ reaction has been 
performed at the storage ring COSY  
in the Research Center J\"ulich in Germany.
In this article we present preliminary results obtained at 
excess energy Q~=~10~MeV, where the predictions of the models 
differ at most.

\section{ Experimental method}
\label{sec:experimental}

In the measurements a polarized proton beam~\cite{prasuhn, stockhorst} with the momentum of 
p$_{beam}$~=~2010~MeV/c has been used. This beam momentum value corresponds to the excess energy $Q~=~10$~MeV. 
The beam of protons has been scattered on the $H_2$ clusters, 
which were produced by the hydrogen cluster target~\cite{dombrowski}.
The COSY-11 detection setup~\cite{brauksiepe,jurek} has been used for the detection of the reaction products.  

\begin{figure}[h]
\begin{center}
\includegraphics[width=4.5in]{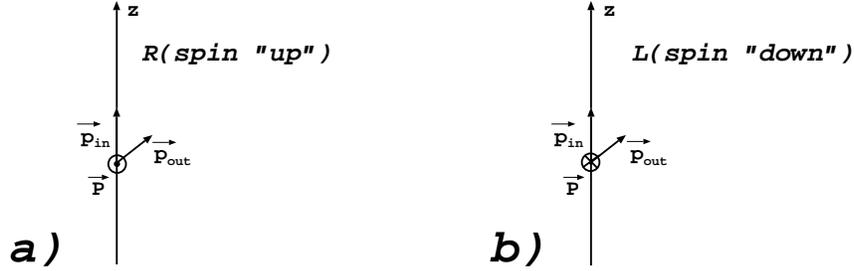} \\
\end{center}
\caption{%
Definition of the scattering to the right (a) and to the left side (b) with respect to the polarization
plane. $\vec{P}$ is a polarization vector of the beam of protons, 
whereas $\vec{p}_{in}$ and $\vec{p}_{out}$ denote the momenta of incoming and outgoing 
particles, respectively. The polarization vector of Fig.~\ref{fig:vectors}.a points 
out of the paper, whereas the polarization vector of Fig.~\ref{fig:vectors}.b points into the paper. 
%The corresponding momenta are equal to $\vec{p}_{in}=\hbar \vec{k}_{in}$ and
%$\vec{p}_{out}=\hbar \vec{k}_{out}$. 
}
\label{fig:vectors}
\end{figure}

As the COSY-11 system is a one-arm detection setup, in order to register 
both types of events: scattered to the right and to the left side (with respect to the  
polarization plane) there is a need to perform the experiment with the separate cycles, in 
which the spin of the proton beam has to be flipped, 
as schematically shown in Fig.~\ref{fig:vectors}. 
In this figure we also define what is meant by 
the terms: "scattering to the right" and "to the left". This definition 
strictly follows the Madison convention~\cite{XXX}.
%\begin{figure}[t]

\begin{figure}[H]
\begin{center}
\includegraphics[width=3.5in]{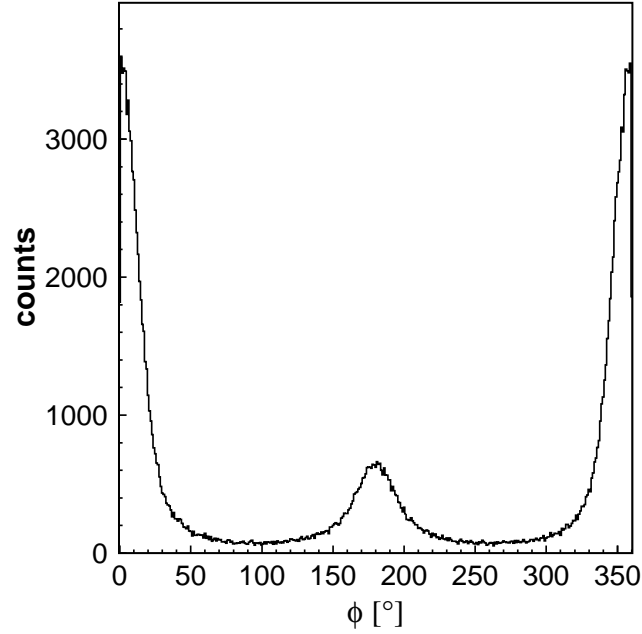}
\end{center}
\vspace{-0.5cm}
\caption{% 
Monte-Carlo simulation of the distribution of the events as 
a function of the angle $\phi$. 
}
\label{katfi}
\end{figure}

In the following we define the polarization plane as a plane spanned by the beam momentum
vector $\vec{p}_{beam}\equiv [0,0,p^z_{beam}]$ and a polarization vector
$\vec{P}=[0,P,0]$.
%$\hat{n}=\frac{\vec{k}_{in} \times \vec{k}_{out}}{|\vec{k}_{in} \times \vec{k}_{out}|}$
%vectors.
If the incident beam consists of spin $\frac{1}{2}$ particles with 
a transverse polarization, which was the case in our experiment, 
the formula for the differential cross section $\sigma(\theta,\phi)$ for a scattering into the 
solid angle around $\theta$ and $\phi$ angles reads:  
\begin{equation}
\sigma(\theta,\phi) = \sigma_0(\theta) (1+A_y(\theta)\vec{P} \cdot \hat{n}),
\label{eq:cross}
\end{equation}
%WAZNE!!!
%do pracy doktorskiej dopisac, ze jest to szczegolny przypadek 
%postaci obserwabli polaryzacyjnych, ktore zostaly zdefiniowane 
%w paragrafie 5 pozycji G.G.Ohlsen, Rep. Prog. Phys 1972, 35, 717-801. 
where $\sigma_0(\theta)$ denotes the differential cross section for a scattering
of an unpolarized beam, A$_y$($\theta$) is an analyzing power of the reaction, 
%$\vec{P}$ is a beam polarization vector 
and $\hat{n}$ is a unit vector along 
$\vec{p}_{in} \times \vec{p}_{out}$. $\phi$ is an angle between $\vec{P}$
and $\hat{n}$, i.e. $\vec{P} \cdot \hat{n} = P \cos \phi \sin \theta$ and $\theta$ 
denotes the angle between $\vec{p}_{in}$ and $\vec{p}_{out}$. In the experiment we 
were restricted to the plane corresponding to $\phi=0^\mathrm{o}$ in case of 
spin "down" (or $\phi=180^\mathrm{o}$ for spin "up"), where
the detection efficiency of the COSY-11 setup is the highest and it decreases
drastically when going outside this particular scattering plane (see Fig.~\ref{katfi}). Therefore, in 
what follows we will consider the scattering in this plane solely.  

The total production rate to the right N$_R$ and to the left side N$_L$ 
%into 
%the solid angle around $\theta$ and $\phi$ 
can be expressed as follows: 
\begin{equation}
N_R(\theta) = N(\theta,\phi=\pi) = \sigma_R(\theta,\pi) E_R(\theta,\pi) \int{L_R dt_R} = \sigma_0(\theta) (1-A_y(\theta) P_R) E_R(\theta,\pi) \int{L_R dt_R},
\label{eq:crossR}
\end{equation}
\begin{equation}
N_L(\theta)= N(\theta,\phi=0) = \sigma_L(\theta,0) E_L(\theta,0) \int{L_L dt_L} = \sigma_0(\theta) (1+A_y(\theta) P_L) E_L(\theta,0) \int{L_L dt_L},
\label{eq:crossL}
\end{equation}
where the E$_R(\theta,\pi)\equiv$~E$_L(\theta,0)$ is the function of the efficiency of the COSY-11 detection system, 
L$_{R/L}$ stands for the luminosity for spin up/down, and t$_{R/L}$ denotes the time of the measurement 
with spin up/down, respectively. The polarization degrees P$_R$ for spin up and P$_L$ for spin down are equal
within the 5\% of accuracy, as has been shown in the previous measurements by means of the
EDDA polarimeter~\cite{altmeier}. Therefore for the further considerations we will assume 
that P$_L$=P$_R\equiv$~P. Dividing equation~\ref{eq:crossL} by~\ref{eq:crossR}, and 
introducing a relative luminosity:
\begin{equation} 
L_{rel} \equiv \frac{\int{L_R dt_R}}{\int{L_L dt_L}},
\label{relative}
\end{equation} 
results in the following formula for the analyzing power: 
\begin{equation}
A_{y}(\cos\theta)=\frac{1}{P} \frac{L_{rel} N_L(\cos\theta) - N_R(\cos\theta)}{L_{rel} N_L(\cos\theta) + N_R(\cos\theta)}, 
\label{eq:analysing}
\end{equation}

Therefore, the values of the relative luminosity L$_{rel}$, the spin-averaged polarization degree P, 
together with the production rates to the right N$_R$, and to the left side N$_L$ 
are the physical observables to be found in order to 
determine the analyzing power function A$_y$. In the following subsections the methods
used to obtain these quantities will be presented.

\subsection{ Calculations of the relative luminosity and polarization}
\label{subsec:polarisation}

The experiment has been performed with the 
300 s long cycles, for which the spin of the incident proton beam has been
flipped from cycle to cycle. This method was intended to reduce the systematical uncertainties
caused by the change of the beam parameters due to the variation of the target densities. 
The duration of the cycle has been
set up in such a manner, that it was significantly shorter than the time
scale (10 hours) for the substantial changes of the density of the target.

For the determination of the relative luminosity a detection system 
consisting of two pairs of the scintillators placed in the 
polarization plane have been used~\cite{czyzyk}. Due to the parity invariance, 
a cross section for any scattering into this plane does not depend on the polarization degree. 
Thus, the number of the coincidences $n$ between scintillators 
has been used as a measure of the absolute luminosity.
Following the definition given in eq.~\ref{relative}, we can express the relative luminosity 
as a ratio of the number of coincidences in the polarization plane 
during the cycles with spin "up" ($n_R$) and "down" ($n_L$):
\begin{equation}
L_{rel} = \frac{n_R}{n_L}.
\label{gowienko}
\end{equation}
The relative luminosity determined using eq.~\ref{gowienko}
was found to be L$_{rel}$=0.96468$\pm$ 0.00065.

\begin{figure}[H]
\begin{minipage}[t]{0.48\linewidth}
\begin{center}
\includegraphics[width=2.5in]{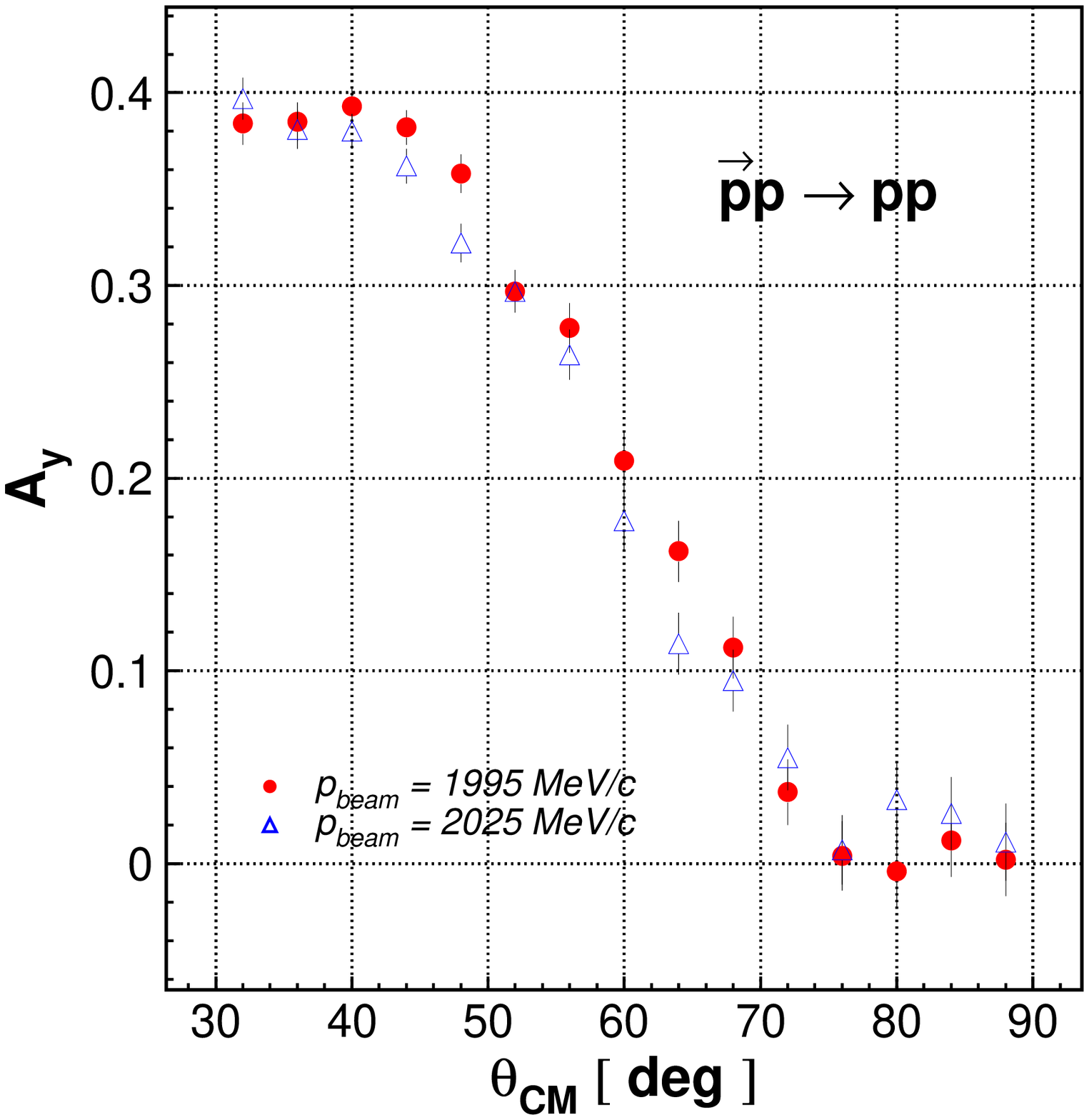} \\
\end{center}
\caption{Analyzing powers for the $\vec{p}p\to pp$ elastic scattering as determined 
by the EDDA collaboration~\cite{altmeier} at the beam momenta specified inside a figure. 
To obtain the analyzing powers for a beam momentum used in the 
experiment a linear approximation has been applied.  
}
 \label{edda}
\end{minipage}%
\hspace{0.04\textwidth}%
\begin{minipage}[t]{0.48\linewidth}
\begin{center}
\includegraphics[width=2.5in]{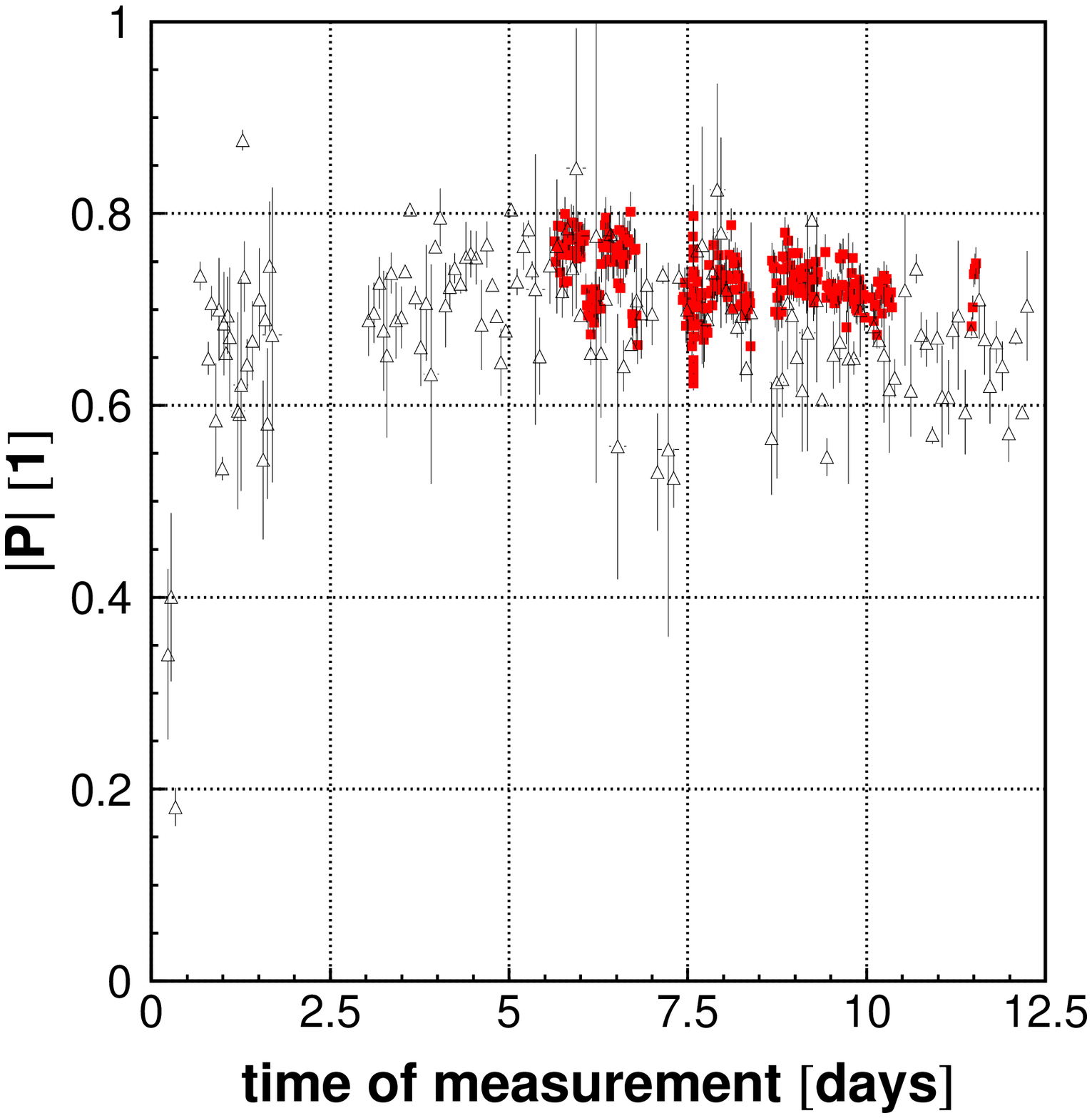} \\
\end{center}
 \caption{ Average polarization degree versus the time of the measurement. 
Open circles are the results obtained by means of the COSY polarimeter, whereas
the full dots are the polarization values determined using the method described in~\cite{czyzyk}.
Data obtained using the independent methods are in line. 
}\label{polar}
\end{minipage}
\end{figure}

The detailed method of determination of the averaged polarization value 
has been described in~\cite{czyzyk}. For the calculations the assymetry of the elastically scattered 
events in the plane perpendicular to the polarization plane has been found for different 
scattering angles in the center-of-mass system (CM). In order to determine the values of the analyzing powers
for the beam momentum of p$_{beam}$~=~2010~MeV, the linear interpolations between available experimental results 
%~\footnote{There 
%is no measurement of the analyzing power for pp scattering at the beam momentum 
%$p_{beam}=2010 MeV/c$, therefore the linear interpolation of the v }
for the proton-proton elastic scattering at $p_{beam}~=~1995$ and 2025~MeV/c have been used. 
Two data sets of the analyzing powers for the $\vec{p}p\to pp$ reaction for the 
mentioned above beam momenta,
%incident beam momenta lying closest 
%to the beam momentum used in our experiment, 
as measured by the EDDA collaboration~\cite{altmeier} at different CM scattering angles, 
are presented in Fig.~\ref{edda}. 

Fig.~\ref{polar} depicts  
the variation of the beam polarization during the time of the measurement. Data sets 
obtained using the COSY polarimeter~\cite{bauer}(open circles) are confronted with the results of the measurements with 
the COSY-11 polarimeter~\cite{isabela,kowina,czyzyk}(full circles).  
An agreement between both sets of data is visible in Fig.~\ref{polar}. 
The average value of the beam polarization for the whole period 
of the measurement was found to be P=0.680$\pm$ 0.010.

\subsection{ Determination of the background free production rates for the $\vec{p}p\to pp\eta$ reaction}
\label{subsec:rl}

\begin{figure}[h]
\begin{center}
\includegraphics[width=2.5in]{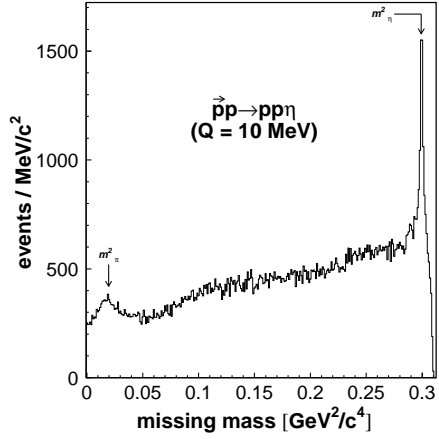} \\
\end{center}
\caption{ Spin-averaged missing mass spectrum for the $\vec{p}p\to pp\eta$ reaction at the 
excess energy Q~=~10 MeV, as measured by means of the COSY-11 detector setup.  
}
\label{missmass}
\end{figure}

A typical missing mass spectrum for the $\vec{p}p\to pp\eta$ reaction as measured using the 
COSY-11 detection setup is presented in Fig.~\ref{missmass}. Figure shows the spin-averaged missing mass spectrum, gathered
during the whole time of the experiment. Over the wide multi-pionic background a clear peak
originating from the $\eta$ meson production is visible.   

Optimizing the statistics and the expected shape of the analyzing power function, 
the range of the center-of-mass polar angles of the $\eta$ meson emission  
has been divided into four bins. 
To separate the actual production rates from the 
background, the reactions with multi-pionic production as well as the events with $\eta$ production have been simulated, using a 
program based on the GEANT3~\cite{geant} code. 
%The response of the 
%COSY-11 detectors implemented into the code has been analyzed in the same way as it had been 
%performed for the experimental data. 
Generated events which fulfilled conditions equivalent to the experimental trigger have been 
analyzed in the same way as the experimental data.  
In such a way we have obtained the missing mass shapes of the background reactions ($pp\to pp2\pi, 3\pi, 4\pi$) 
as well as the shape of the signal ($pp\to pp\eta$). 
In order to perform credible simulations of the missing mass spectra 
the geometry of the drift chambers (used for the momentum reconstruction) as well as the position 
and geometrical parameters of the target have to be known precisely. 
The angle and the relative position of the COSY-11 drift chambers (2 parameters), 
a target position (3 parameters) and a relative shift between the beam and a target (1 parameter)
have been optimized using the elastically scattered events and the 
method described in~\cite{NIM,hab}. 

\begin{figure}[H]
\begin{center}
\includegraphics[width=3.7in]{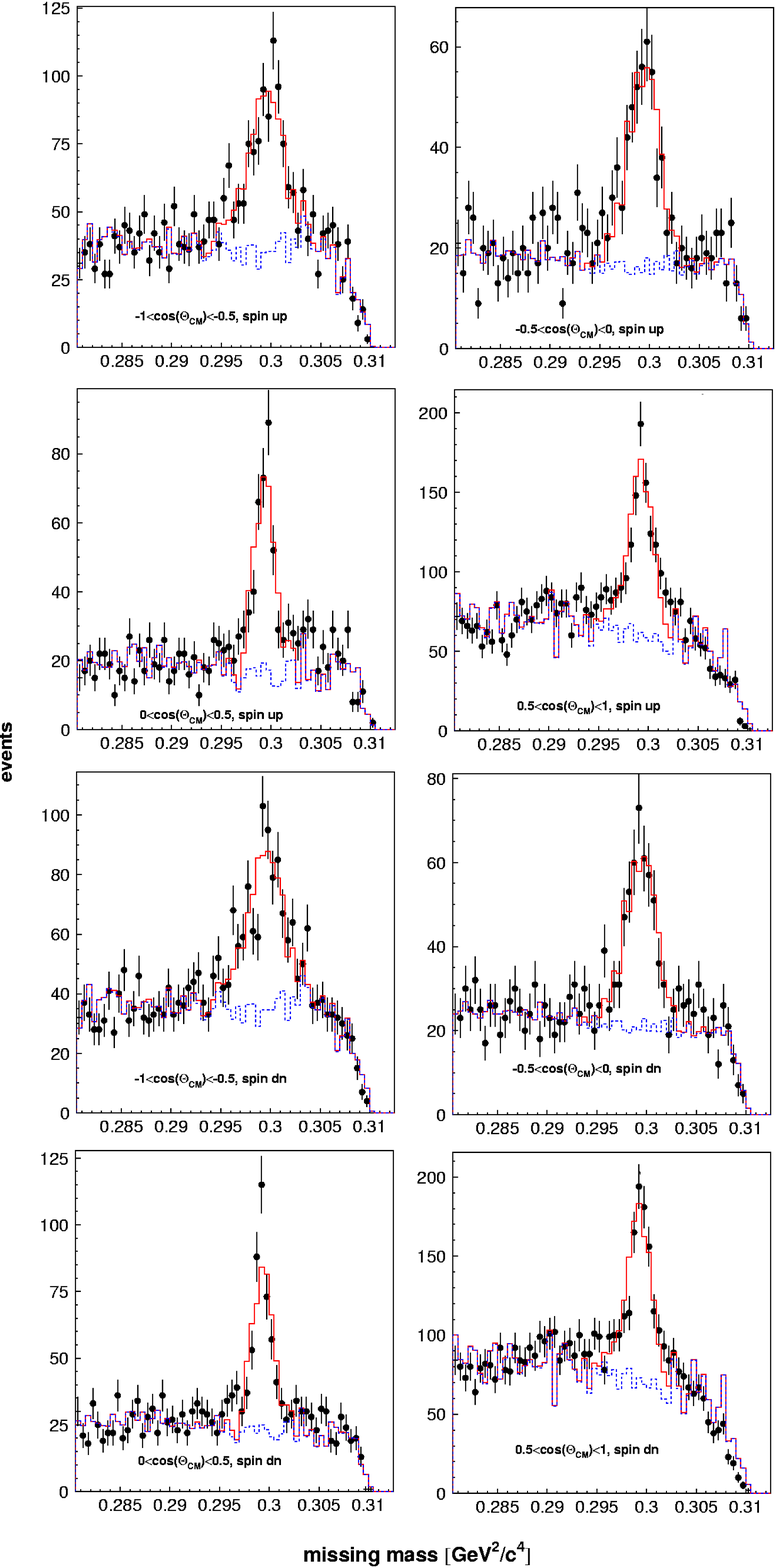} 
\end{center}
\vspace{-1.0cm}
\caption{ Missing mass spectra for different ranges of $\cos\theta_{CM}$ 
for spin up (two upper panels) and spin down (two lower panels).
Dots represent the experimental points. Dotted line shows the generated multi-pionic
background. The full line is a best fit of the sum of the signal and background to the experimental data.
}
\label{up_and_down}
\end{figure}

The multi-dimensional fit of the
simulated missing mass spectra to the corresponding experimental histograms has been performed
using the MINUIT~\cite{minuit} minimization package.
The fit has been performed simultaneously to 8 histograms (see Fig.~\ref{up_and_down}),
each with 60 points.
The $\chi^2$ of the fit have been minimized as a 
function of six parameters: scaling factors of the generated background and signal reactions
(altogether 4 parameters) and 2 parameters responsible for the spread and the absolute 
value of the beam momentum.
%There were 12 free parameters 
%we accounted for in the fit, namely: the scaling factors of the generated background
%and signal reactions (altogether 4 parameters); 2 parameters responsible for the track 
%reconstruction (the angle and the relative position of the COSY-11 drift chambers); 
%target position (3 parameters); relative shift between the beam and a target;  
%%(which has to 
%%be taken into account in the simulation code); 
%and 2 parameters responsible for the 
%shape of the beam and its position on a momentum axis.
The detailed process of the search for the best values of these parameters will 
be described elsewhere~\cite{czyzyk-phd}, here we would only like to 
report that the minimum value of $\chi^2$ was found to be 1.62,
corresponding to the spread of the beam momentum equal to 0.2~MeV/c
and the shift from the nominal beam momentum equal to 2.1~MeV/c. 
These values are in a good agreement with results of previous COSY-11 experiments~\cite{NIM}.

%present a two-dimensional projection of the $\chi^2$ onto the 
%the subspace spanned by the last two parameters mentioned in the list above, where
%the other parameters had been previously fixed. 
%This projection is depicted in Fig.~\ref{chi2}, where the reduced $\chi^2$
%is plotted as a function of two fit parameters: $\sigma$ --- a parameter describing the 
%momentum dispersion of the proton beam, and $\Delta p$ --- an offset 
%of a beam momentum from its nominal position 
%(the Monte Carlo calculations were performed with a beam momenta $p_{beam} = (2010 + \Delta p)$~MeV/c).     
%In the case of this projection, a valley with a minimum at the values
%$\sigma = 0.2$~MeV/c and $\Delta p = 2.1$~MeV/c can be seen. These values were 
%chosen as the best parameters of the fit.  
%\begin{figure}[H]
%\begin{center}
%\includegraphics[width=4.0in]{chi_2.ps} \\
%\end{center}
%\caption{ A projection of a reduced $\chi^2$ of the multi-dimensional 
%fit onto the subspace spanned by $\sigma$ and $\Delta p$. 
%}
%\label{chi2}
%\end{figure}

In Fig.~\ref{up_and_down} the missing masses for the individual 
$\theta_{CM}$ subranges, separately for spin up (upper panel) and down (lower panel), are shown.
Full dots correspond to the experimental data,  
the dotted line depicts the shape of the background, whereas the solid line
represents the best fit of the Monte-Carlo data to the experimental spectra.  

\section{ Results}
\label{results}
%The production rates as the function 
%of $\cos\theta_{CM}$, obtained as the result of the best fit, along with the analyzing powers, calculated according to the
%eq.~\ref{eq:analysing} are shown in Tab.~\ref{analysing-table}. 
%
%\begin{table}[H]
%\begin{center}
%\begin{tabular}
%{|c|c|c|c|}\hline
%$\cos\theta_{CM}$ & {\bf N$_R$} & {\bf N$_L$} & {\bf A$_y$} \\\hline
%[-1;-0.5) & 555 $\pm$ 44 & 500 $\pm$ 42 & -0.103 $\pm$ 0.086 \\\hline 
%[-0.5;0) & 329 $\pm$ 28 & 335 $\pm$ 29 & -0.013 $\pm$ 0.089  \\\hline 
%[0;0.5) & 314 $\pm$ 26 & 343 $\pm$ 29 & 0.038 $\pm$ 0.087    \\\hline 
%[0.5;1] & 717 $\pm$ 49 & 727 $\pm$ 52 & -0.016 $\pm$ 0.073   \\\hline 
%\end{tabular}
%\caption{ Number of events scattered to the individual $\theta_{CM}$ angle and
%corresponding analyzing powers for the $\vec{p}p\to pp\eta$ reaction at the excess energy
%Q=10 MeV. 
%}\label{analysing-table}
%\end{center}
%\end{table}

\begin{figure}[h]
\begin{center}
\includegraphics[width=2.5in]{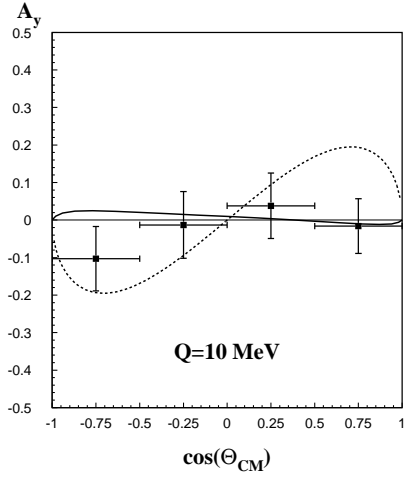} \\
\end{center}
\caption{ Preliminary analyzing power function for 
the $\vec{p}p\to pp\eta$ reaction at Q=10~MeV. Vertical 
bars denote the statistical error, whereas the horizontal bars 
stand for the ranges of averaging.  
}
\label{fig-analysing}
\end{figure}

Preliminary results of the analyzing power function A$_y(\cos\theta)$ 
are presented in Fig.~\ref{fig-analysing} as the full dots. The dotted 
line shows the predictions determined according 
to the vector meson exchange model~\cite{wilkin}, whereas the solid 
line refers to the pseudoscalar meson exchange model~\cite{nakayama}.
The errors are of the statistical nature only. 

\section{ Conclusions}
\label{conclusions}
The preliminary analyzing power results are bared with rather large 
statistical uncertainties, therefore a statement on
the mechanism of the $\eta$ meson production relying on this set of data is 
currently unclear.
As can be seen from Fig.~\ref{fig-analysing} the data point at 
$\cos \theta = 0.75$
tend to prefer 
the pseudoscalar meson exchange model~\cite{nakayama},
whereas the
data point at $\cos \theta = -0.75$ is slightly more in line 
with the predictions of the vector meson exchange model.   
The results show rather small values of the analyzing power
in the close-to-threshold region, which may be the indication 
of the $\eta$ production to the $s$-wave final state, solely.
However, for this statement the exact partial wave 
analysis remains to be done.

\begin{ack}
%The work has been supported by the European Community - Access to
%Research Infrastructure action of the Improving Human Potential Programme,
%by the FFE grants (41266606 and 41266654) from the Research Centre
%J{\"u}lich, by the DAAD Exchange Programme (PPP-Polen),
%by the Polish State Committe for Scientific Research
%(grant No. PB1060/P03/2004/26), and by the RII3/CT/2004/506078
%- Hadron Physics-Activity -N4:EtaMesonNet.
We acknowledge the support of the
European Community-Research Infrastructure Activity
under the FP6 "Structuring the European Research Area" programme
(HadronPhysics, contract number RII3-CT-2004-506078),
of the FFE grants (41266606 and 41266654) from the Research Centre J{\"u}lich,
of the DAAD Exchange Programme (PPP-Polen),
of the Polish State Committe for Scientific Research
(grant No. PB1060/P03/2004/26), \\
and of the
RII3/CT/2004/506078 - Hadron Physics-Activity -N4:EtaMesonNet.
\end{ack}

\end{document}